\documentclass[11pt]{article}
\usepackage{jcappub,amsmath,amssymb,hyperref,framed}
\usepackage{mathtools,mathrsfs}
\usepackage{bm}
\usepackage{amsmath}
\usepackage{amssymb}
\usepackage{bbold}
\usepackage{pgfplots}
\usepackage{graphics}
\usepackage{todonotes}
\usepackage{tikz}
\usetikzlibrary{arrows,positioning,decorations.markings,decorations.pathmorphing,calc}
 
\begin{document}

\title{Gravitational couplings in Chameleon models}

\author[1]{Macarena Lagos,}
\author[2]{Hanjue Zhu}
%\author[3]{Wayne Hu}

\def\hj#1{\textcolor{cyan}{HJ: #1}}

\affiliation[1]{Kavli Institute for Cosmological Physics, The University of Chicago, Chicago, IL 60637, USA} 
\affiliation[2]{The University of Chicago, Chicago, IL 60637, USA} 
%\affiliation[3]{Kavli Institute for Cosmological Physics, Department of Astronomy \& Astrophysics, Enrico Fermi Institute, The University of Chicago, Chicago, IL 60637, USA}

\emailAdd{mlagos@kicp.uchicago.edu}
\emailAdd{hanjuezhu@uchicago.edu}
%\emailAdd{whu@background.uchicago.edu}

\abstract{
We consider cosmological models where dark energy is described by a dynamical field equipped with the Chameleon screening mechanism, which serves to hide its effects in local dense regions and to conform to Solar System observations. In these models, there is no universal gravitational coupling and here we study the effective couplings that determine the force between massive objects, $G_N$, and the propagation of gravitational waves, $G_{gw}$. In particular, we revisit the Chameleon screening mechanism without neglecting the time dependence of the galactic environment where local regions are embedded in, and analyze the induced time evolution on $G_N$ and $G_{gw}$, which can be tested with Lunar Laser Ranging and direct gravitational waves observations. 
We explicitly show how and why these two couplings generically differ.
We also find that due to the particular way the Chameleon screening mechanism works, their time evolutions are highly suppressed in the weak-field non-relativistic approximation.
}
\keywords{Dark Energy, Screening Mechanism, Gravitational Waves}

\maketitle
%\newpage

\setcounter{tocdepth}{2}
%\tableofcontents

\section{Introduction}

The origin to the observed late-time accelerated expansion of the Universe still remains one of the biggest challenges in cosmology \cite{Riess:1998cb}. The concordance $\Lambda$CDM model assumes this expansion is driven by the presence of a dark energy component with a constant energy density (a cosmological constant), which makes up nearly $70\%$ of the total energy density of the universe today \cite{Aghanim:2018eyx}. In order to conform to cosmological observations, the value of the cosmological constant must be incredibly small. From particle physics arguments we can predict the existence of vacuum energy, which provides a value for the cosmological constant, but current estimates imply that it is more than 50 orders of magnitude larger than the observed value \cite{Martin:2012bt}. This constitutes the so-called cosmological constant problem and shows how the $\Lambda$CDM model is at odds with well-established and robust particle physics theory. This problem has motivated the study of alternative models where dark energy can change in time and where there is a dynamical evolution that drives its energy density to a small value today (see e.g.~\cite{Jain:2010ka, Clifton:2011jh, Joyce:2014kja, Koyama:2015vza, Joyce:2016vqv, Nojiri:2017ncd, Frusciante:2019xia} for reviews).

In the simplest alternative models to $\Lambda$CDM, dark energy is described by a dynamical scalar field mediating a long-range force that affects cosmological scales. If the dark energy field has non-trivial interactions with gravity, it may also a priori affect shorter scales, such as Solar System scales. However, since gravity is well tested within the Solar System, these alternative models are necessarily equipped with a mechanism that hides the dark energy from local observations. This is achieved through screening mechanisms, which rely on the high density or small scales of local environments (relative to the cosmological density and scale) to suppress the fifth force mediated by the dark energy field. The typical mechanisms considered in the literature rely on the dark energy field acquiring a weak coupling to matter, a large mass, or large inertia in local regions (see a review in \cite{Jain:2010ka, Joyce:2014kja}). In this paper, we focus on the Chameleon screening mechanism \cite{Khoury_2004, Khoury:2003rn}, in which the dark energy field has a mass that depends on its environment. In dense environments, the Compton wavelength of this field is small, and hence it mediates an effectively undetectable short-range force. On the contrary, around diffuse environments, the scalar field becomes light with a large Compton wavelength, and potentially detectable on cosmological scales. 

Furthermore, the presence of non-trivial interactions may break the equivalence principle \cite{Hui:2009kc}, in which case the strength of interactions will depend on the specific bodies at play and there will not be a universal gravitational coupling anymore. In particular, two relevant couplings corresponding to $G_N$ (determining the matter-matter interactions of gravity) and $G_{gw}$ (determining the self-interactions of gravity) are expected to be generically different \cite{Tahura:2018zuq,Wolf:2019hun}. Meanwhile, in the $\Lambda$CDM model, both of these couplings coincide and are constants, given by Newton's gravitational constant $G$. The presence of a dynamical dark energy may then affect observations in different ways. Observations from planetary motion will be mainly concerned with testing $G_N$, whereas gravitational wave (GW) observations with $G_{gw}$. 

Screening mechanisms are usually studied in an approximately static local weak-field regime of gravity. However, when the dark energy field has a non-trivial interaction with gravity, its time evolution can leak into the short scales and induce a time evolution of the gravitational force in local regions such as the Solar System. Indeed, this was shown to be the case for some models with Vainshtein screening \cite{Babichev:2011iz, Kimura:2011dc} where it was found that the Newtonian gravitational coupling $G_N$ can have a time variation on Hubble timescales. However, the time variation of $G_N$ has been constrained with observations from Lunar Laser Ranging (LLR) \cite{Hofmann_2018}, where it has been found that $|\dot{G}_N/G_N|\lesssim  10^{-3}H_0$, with $H_0$ the Hubble rate today. 

Similarly, it has been noticed that dynamical dark energy may also induce a time variation in the coupling $G_{gw}$. This time variation can be measured with multi-messenger detections of both electromagnetic (EM) and GW signals, through comparisons of the EM luminosity distance $d^{em}_{L}$ and the so-called GW luminosity distance $d^{gw}_{L}$. The latter probes the decay of the GW amplitude as it travels over cosmological distances. Specifically, these distances are related by \cite{Deffayet:2007kf, Saltas:2014dha, Lombriser:2015sxa, Amendola:2017ovw, Lagos:2019kds, Belgacem:2019pkk, Dalang:2019rke, Garoffolo:2019mna}: 
\begin{equation}\label{dgw}
d^{gw}_{L}(z_s)/d^{em}_{L}(z_s)  =\sqrt{G_{gw}(z_s,r_s)/G_{gw}(z_o,r_o)},
\end{equation}
where $z_s$ and $z_o$ are the redshifts of the source and observer, respectively. In \cite{Amendola:2017ovw, Dalang:2019fma} it has been argued and clarified that what GW probes is the difference in the {\it local} value of $G_{gw}$ at the moment of emission and detection (as opposed to its cosmological value), so observations depend also on the positions of the source, $r_s$, and observer, $r_o$. From here we see that by  probing the observables on the LHS of this equation we would provide constraints on the time evolution of $G_{gw}$.
Recent forecasts show that GW observations could impose bounds that range between $|\dot{G}/G|\lesssim  \mathcal{O}(1)H_0$ for LIGO \cite{Lagos:2019kds}, and $|\dot{G}/G|\lesssim  \mathcal{O}(10^{-2})H_0$ for  LISA \cite{Belgacem:2019pkk} (and similar bounds for other detectors such as Einstein Telescope, DECIGO, Voyager, and Cosmic Explorer \cite{Nishizawa:2019rra, DAgostino:2019hvh, Bonilla:2019mbm}). 

In this paper, we consider dark energy models with Chameleon screening and obtain the explicit time evolution of both $G_N$ and $G_{gw}$ in the weak-field regime, and discuss whether they conform to observations. Chameleon screening works in such a way that it is sensitive to the galactic background in which the local screened system is embedded. Therefore, since time variations of the background can be even faster than cosmological timescales, these models will only conform to current LLR constraints if screening suppresses this background time evolution (contrary to models with Vainshtein screening, as previously mentioned). 
Other phenomenological aspects of Chameleon models have been extensively studied (see e.g.~\cite{Burrage:2016bwy} for a review). It has been found that they  cannot simultaneously screen and self-accelerate cosmologically \cite{Wang:2012kj} (i.e.~as in the late-time acceleration of the Universe is completely driven by the non-trivial interactions between gravity and the scalar field). But they can still act as a negative pressure component in the Friedmann cosmological equation \cite{Copeland:2006wr}, similar to the cosmological constant. Also, the effect of the scalar field must be suppressed on cosmological scales in order to avoid significant enhancement on the growth of large-scale structures \cite{Song:2006ej,Hu:2007nk, Tsujikawa:2007gd, 2009PhRvD..80h3505S}.
Furthermore, Chameleon models belong to the family of theories with a luminal propagation speed of GW \cite{Baker:2017hug, Langlois:2017dyl, Creminelli:2017sry, Sakstein:2017xjx, Ezquiaga:2017ekz} and thus they conform to the latest constraints from time delays of EM and GW signals from the event GW170817 \cite{2041-8205-848-2-L13, 2041-8205-848-2-L12}.
Searches for new tests are still ongoing \cite{Naik:2019moz, Pernot-Borras:2019gqs, Arnold:2019vpg, Arnold:2019zup}.

In this paper, we first confirm that both $G_N$ and $G_{gw}$ couplings are technically different in Chameleon models. We find that this happens because $G_N$ describes the matter-matter interactions and therefore it receives a direct contribution from the dark energy fifth force, whereas $G_{gw}$ does not.
However, if the fifth force is highly screened in local regions, both of these couplings coincide. Regardless, the time evolution of the galactic background is found to be highly suppressed and therefore the time variations induced by dark energy in the couplings $G_N$ and $G_{gw}$ will lie below the sensitivity of LLR and GW observations for these models.
Nevertheless, this model provides a simple example to explicitly illustrate the fact that gravitational couplings are not universal when dark energy has non-minimal interactions with gravity, and that these couplings do evolve in time. More complicated models exhibiting a different screening mechanism, such as Degenerate Higher-Order Scalar-Tensor (DHOST) theories \cite{Langlois:2017mxy}, also predict different $G_N$ and $G_{gw}$ \cite{Crisostomi:2017pjs} and may contain enough freedom to have a suppressed time varying $G_N$ to conform to LLR data, while still allowing for a $G_{gw}$ varying on cosmological timescales. Models like this are potentially falsifiable with future GW detections. 
 
This paper is structured as follows. In Section \ref{Sec:Model} we review the Chameleon model and its equations of motion. In Section \ref{Sec:screen} we obtain the solutions of screening around a time-dependent background and calculate both couplings $G_N$ and $G_{gw}$ and their time evolutions. Finally, in Section \ref{Sec:discussion} we summarize our results and discuss their consequences. We will be using a mostly positive signature for the metric and a unity speed of light and Planck constant $c=\hbar=1$. 

%--------------------------------------------------------------------------------------------------------------------------------------------------------

\section{Chameleon Model}\label{Sec:Model}
Chameleon scalar fields mediate a fifth force between massive bodies, with a range that decreases with increasing ambient matter density, thereby avoiding its detection in regions of high density  (see a review in \cite{Burrage:2017qrf}). This can be achieved with the following action in the Einstein frame:
\begin{equation}\label{Action}
	S=\int d^4x\, \left[ \sqrt{-g}\left(\frac{M_P^2}{2}R- \frac{1}{2}\varphi_{,\mu}\varphi^{,\mu}-V(\varphi) \right) -  \mathcal{L}_m(A^2(\varphi)g_{\mu\nu},f_m)\right],
\end{equation}
where $g_{\mu\nu}$ is the metric in this frame, $R$ the Ricci scalar, $M_P$ the constant Planck mass, $f_m$ a proxy for any matter field, and $A(\varphi)$ some function describing the non-trivial interaction between the dark energy scalar field $\varphi$ and matter. Typically, this function is considered to be:
\begin{equation}
	A^2(\varphi)= e^{2\beta \varphi/M_P},
\end{equation}
where $\beta$ is some arbitrary constant. 
Also, the potential $V(\varphi)$ is typically considered to be a power law:
\begin{equation}\label{Potential}
	V(\varphi)= \frac{M^{4+n}}{\varphi^n},
\end{equation}
where $M$ is an arbitrary constant with units of mass and $n$>$-1$ or an even negative integer (in order to have a potential that can lead to a Chameleon screening). 
A well-known theory belonging to the class of Chameleon models is the $f(R)$ theory (see \cite{DeFelice:2010aj} for a review), where the Einstein-Hilbert action is allowed to be a generic function of the Ricci scalar $R$. This theory can be recast in a scalar-tensor form such that it looks like eq.~(\ref{Action}).

The equations of motion of this action are given by:
\begin{align}
&\Box\varphi= V_{,\varphi}-\frac{\beta}{M_P}T_{m}\equiv V_{\text{eff},\varphi},\\
&G_{\mu\nu}=R_{\mu\nu}-\frac{1}{2}R g_{\mu\nu}= M_P^{-2}(T_{\mu\nu\varphi} +T_{\mu\nu m}),
\end{align}
where the subscript with colons denote derivatives, and we have defined an effective potential $V_\text{eff}(\varphi,\rho)$ that must have a minimum for Chameleon screening to work successfully. We have also introduced the stress-energy tensors of matter and scalar field:
\begin{align}
&T_{\mu\nu m}=\frac{2}{\sqrt{-g}}\frac{\delta \mathcal{L}_{m}}{\delta g^{\mu\nu}}\label{EQphi},\\
	& T_{\mu\nu \varphi} = \partial_\mu \varphi \partial_\nu\varphi -g_{\mu\nu}\left(\frac{1}{2} \partial_\mu \varphi\partial^\mu \varphi+ V(\varphi) \right)\label{EQg},
\end{align}
with $T_m=g^{\mu\nu}T_{\mu\nu m}$ being the trace.
We emphasize that $T^{\mu\nu}{}_{m}$ is not covariantly conserved in the Einstein frame.
Indeed, from the equations of motion we have that
\begin{equation}
\nabla^\mu T_{\mu\nu m}= -\nabla^\mu T_{\mu\nu \varphi}= -(\Box\varphi-V_{,\varphi})\nabla_\nu\varphi= \frac{\beta}{M_P}T_m\nabla_\nu\varphi.
\end{equation}

We note that we can always perform a transformation of the action (\ref{Action}) such that matter is minimally coupled to the metric. This is the Jordan frame, in which test particles follow geodesics of the effective metric $g_{\text{eff}\mu\nu}=A^2(\varphi)g_{\mu\nu}$ and therefore this is the metric that describes the standard concept of spacetime. For this reason, astronomical observations will be concerned with  $g_{\text{eff}\mu\nu}$ instead of $g_{\mu\nu}$. Nevertheless, it is sometimes easier to work in the Einstein frame. In the Jordan frame, 
the stress energy tensor is indeed conserved: $\nabla^\mu_\text{eff} T_{\text{eff} m}^{\mu\nu}=0$ and therefore we will consider $T_{\text{eff} m}^{\mu\nu}$ to be the external quantity determined by the desired physical setting. For a pressureless fluid, we will have that $\rho_\text{eff}=-g^{\mu\nu}_{\text{eff}}T_{\text{eff}\mu\nu m}=A^{-4}\rho$, with $\rho$ being the energy density in the Einstein frame.

%--------------------------------------------------------------------------------------------------------------------------------------------------------

\section{Time-dependent Screening}\label{Sec:screen}
Since both observations of $G_N$ and $G_{gw}$ depend on the local environment where screening is active, we must obtain the solution of the metric in the screened regime. In addition, since we are interested in tracking their time evolutions, we cannot neglect the time variations of the environment in which the local system is embedded. 

Let us assume we have one source, which could correspond to a planet or star, that is embedded in an environmental background that could correspond to the local neighborhood of a galaxy.
In this case, we consider the following ansatz for the metric in the Einstein frame, corresponding to an expansion of perturbations in the Newtonian gauge around an average homogeneous and isotropic background,
\begin{equation}
ds^2= -(1+2\Phi)d\tau^2+a^2(1-2\Psi)d\vec{x}^2
\end{equation}
where $\Phi(\tau,\vec{x})$ and $\Psi(\tau,\vec{x})$ are much smaller than 1, and describe linear inhomogeneous perturbations due to the presence of the local source. In this case, we can also write perturbatively the scalar field $\varphi$ and the matter energy density $\rho_\text{eff}$ of a pressureless perfect fluid as:
\begin{align}
\varphi&=\varphi_b(\tau)+\delta\varphi(\tau,\vec{x}),\\
\rho_\text{eff}&=\rho_{\text{eff}b}(\tau)+\delta\rho_\text{eff}(\tau,\vec{x}),
\end{align}
where the subscript $b$ refers to the background, which is generically allowed to depend on time here, yet, for simplicity, homogeneous and isotropic on the scales of interest. We emphasize that the non-linear structure of the potential $V$ is crucial when solving the background evolution, as this will allow to obtain a $\varphi_b$ such that the effective mass of the scalar field grows with higher densities $\rho_{\text{eff}b}$ and therefore the fifth force has a short range. 

The background equations are given by:
\begin{align}
&\ddot{\varphi}_b+3H\dot{\varphi}_b=-V_{\text{eff},\varphi}=-V_{,\varphi}(\varphi_b)-\frac{\beta}{M_P}A^4_b\rho_{\text{eff}b},\\
&3H^2M_P^2=A^4_b\rho_{\text{eff}b}+\frac{1}{2}\dot{\varphi}_b^2+V(\varphi_b),
\end{align}
where $A_b=A(\varphi_b)$, which we will require to be close to unity for consistency of the perturbation theory, that is, we will need $|\beta\varphi_b/M_p|\ll1$. Since $V(\varphi)$ is constructed in such a way that $V_{\text{eff}}$ has a minimum, then the background dynamics will be such that $\varphi_b$ evolves towards the minimum of the potential, and will either decay with larger $\rho_{\text{eff}b}$ if $n$>$-1$ or grow if $n<-1$.

Next, we consider the presence of a source on top of the background, and solve for the linear perturbations of the metric and scalar field. The equation of motion for $\varphi$ yields: 
\begin{equation}
g^{\mu\nu b}\partial_\mu \partial_\mu \delta \varphi  + 3 H \dot{ \delta\varphi}  +\dot{\Phi} \dot{ \varphi_{b}} + 3 \dot{\Psi} \dot{ \varphi_{b}} +2\Phi\ddot{\varphi}_b +6(\Phi+2\Psi)H\dot{\varphi}_b= m_{\text{eff}}^2(\varphi_b,\rho_{\text{eff}b})\delta\varphi+ \frac{\beta}{M_P}A_b^4\delta \rho_{\text{eff}},
\end{equation}
where overdots denote derivatives with respect to $\tau$. Here, $m_{\text{eff}}^2= V_{\text{eff},\varphi\varphi} (\varphi_b,\rho_{\text{eff}b})$ corresponds to the effective mass that the perturbed scalar field $\delta\varphi$ acquires due to the background it is embedded in.

For the scenarios in consideration, the background will evolve in a characteristic timescale $T$ such that the spatial variations of the perturbation fields on small scales will be much larger than the time variations set by $T$. For instance, we can consider the background to have galactic time variations of order of Gyr, meanwhile Solar System scales will have spatial variations of order 1AU for the metric potentials, and even smaller for the scalar field that mediates a short-range force.
In these cases, we can make a quasi-static approximation where we neglect all the time derivatives compared to spatial derivatives so that the scalar equation simplifies to: 
\begin{equation}
\nabla^2\delta \varphi \approx  m_{\text{eff}}^2(\varphi_b,\rho_{\text{eff}b})\delta\varphi+ \frac{\beta}{M_P}A_b^4\delta \rho_\text{eff},
\end{equation}
where $\nabla^2=g^{bij}\partial_i\partial_j$. 
If we consider the source $\delta \rho_\text{eff}$ to be given by a spherically symmetric body of total constant mass $M$ and radius $R$, the solution to this equation outside of the source will be given by \cite{Burrage:2017qrf}:
\begin{equation}
\delta \varphi = -2GA_b^2M\frac{\beta}{r}\left(1-\frac{M(r_s)}{M}\right)e^{-m_\text{eff}(r-R)/A_b},
\end{equation}
where $r$ is the physical distance in the Jordan frame, that has been defined as $r=a_\text{eff}\sqrt{\delta^{ij}x^ix^j}$, with $a_\text{eff}=aA_b$ being the Jordan frame scale factor. Here, $G$ is the gravitational constant such that $8\pi G=M_P^{-2}$, and $r_s$ is the screening radius that should be close to $R$ if spatial screening is effective. In this case, the scalar field is sourced only by a very thin shell near the surface of the massive body that suppresses the overall amplitude of the fifth force even if the range of the force is long. Also, the potential $V(\varphi)$ is chosen such that $m_\text{eff} $ increases with density and therefore the range of the fifth force gets shorter in denser environments.

On the other hand, the equations for the metric lead to $\Phi=\Psi$ (there is no anisotropic stress in this frame when considering linear perturbations) and a modified Poisson equation:
\begin{align}
M_P^2\nabla^2 \Phi&=\frac{1}{2}\delta\rho_{\varphi}-\frac{3}{2}H(\rho_{\varphi}+p_{\varphi})v_{\varphi}+ \frac{1}{2}\delta \rho\\
&=
\frac{1}{2}\dot{ \delta\varphi} \dot{\varphi}_b + \frac{1}{2}(V_{,\varphi}+3H\dot{\varphi}_b)\delta\varphi -\frac{1}{2}\Phi\dot{\varphi}_c^2 + \frac{1}{2}\frac{\beta}{M_P}\rho_{\text{eff}b}\delta\varphi 	+ \frac{1}{2}A_b^4\delta \rho_\text{eff} ,
\end{align}
where we have used that $v_{\varphi}=-\delta\varphi/\dot{\varphi}$.
Next, we assume that spatial screening is effective, and therefore the gradient of $\Phi$ is dominated by the source $\delta\rho_\text{eff}$. We thus neglect the contributions by $\delta \rho_{\varphi}$ and similarly for the additional terms that would possibly modify the gradient of the metric potential. In this case, the Poisson equation is simply approximated to:
\begin{equation}
M_P^2\nabla^2\Phi\approx \frac{1}{2}A_b^4\delta\rho_\text{eff},
\end{equation}
from where we see that a time dependence is introduced to the metric potential due to the time evolution of the background. If we consider that the source $\delta\rho_\text{eff}$ is given by a spherically symmetric body, then the metric potential outside the source will be given by:
\begin{equation}
\Phi\approx -\frac{A_b^2GM}{r}.
\end{equation}
When transforming to the Jordan frame, we have that $g_{\mu\nu\text{eff}}=A^2g_{\mu\nu}\approx A^2_b(1+2\beta\delta \varphi)g_{\mu\nu}$ so that:
\begin{align}
&\Phi_\text{eff}\approx \left(\Phi +\frac{\beta}{M_P}\delta\varphi\right),\\
&\Psi_\text{eff}\approx \left(\Psi -\frac{\beta}{M_P}\delta\varphi\right).
\end{align}
Note that the potentials differ from each other in this frame because there is an effective non-vanishing anisotropic stress. 

Test particles move on geodesics of the effective metric (Jordan frame), and this geodesic equation in the weak-field limit has a fifth force per unit mass given by:
\begin{equation}
F_5= -\beta \nabla \delta\varphi/M_P \approx -\frac{GMA_b^2}{r^2} \left(2\beta^2Qe^{-m_\text{eff}/A_b(r-R)}\right),
\end{equation}
where $Q= (1-M(r_s)/M)$. %This means that the extra contribution coming from the scalar field to the metric potential will be given by: $\Delta V= -\beta \delta \varphi/M_P$.
The total gravitational force by a body of mass $M$ is then given by:
\begin{equation}
F_{tot}=- \frac{G_{tot}M}{r^2},
\end{equation}
where
\begin{equation}
G_{tot}=GA_b^2\left[1+2\beta^2Qe^{-m_\text{eff}/A_b(r-R)}\right].
\end{equation}
Note that $G_{tot}$ depends on the body, and hence the weak equivalence principle is broken (not all bodies gravitate equally). Indeed, where $r_s$ lies will depend on the composition of the body, and thus in turn $Q$ will depend on its composition. Therefore, there is no universal gravitational coupling, and if we are interested in the gravitational force between two bodies 1 and 2 with masses $M_1$ and $M_2$, respectively, we will be probing the following force:
\begin{equation}
F_{12}= -G_N\frac{M_1M_2}{r^2},
\end{equation}
where the gravitational coupling is given by
\begin{equation}
G_N=GA_b^2( 1+2\beta^2Q_1Q_2e^{-m_\text{eff}r/A_b}),
\end{equation}
where $r$ is the physical distance between both bodies. 
Note that with data from the Lunar Laser Ranging experiment, we mainly test the force between the Moon and the Earth and therefore we test the time evolution of $G_N$. This time evolution will be given by the following expression in Chameleon models:
\begin{equation}
\frac{\dot{G_N}}{G_N}=    2\frac{\dot{A}_b}{A_b }+\left[\frac{\dot{Q}_1}{Q_1}+ \frac{\dot{Q}_2}{Q_2} -\dot{(m_\text{eff}/A_b)}r\right]\frac{F_{5AB}}{F_{AB}},
\end{equation}
where we have assumed that $Q$ also changes in time as the screening radius is indirectly determined by the background scalar field $\varphi_b(t)$. 

On the other hand, the detected amplitude of gravitational waves will deviate from that expected in GR if there is a local time variation of the GW gravitational coupling $G_{gw}$. 
In Horndeski models (larger family of scalar-tensor theories that Chameleon models belong to), this coupling has been found to be given by the conformal factor $G_{gw}=A^{2}G$ \cite{Dalang:2019rke}, which is clear when looking at the action (\ref{Action}) in the Jordan frame:
\begin{equation}\label{ActionJ}
S=\int d^4x\, \left[ \sqrt{-g_\text{eff}}\left(\frac{M_P^2}{2}A^{-2}R(g_\text{eff})- \frac{1}{2}k^2(\varphi)g^{\text{eff}\mu\nu}\varphi_{,\mu}\varphi_{,\nu}-A^{-4}V(\varphi) \right) -  \mathcal{L}_m(g_{\text{eff}\mu\nu},f_m)\right],
\end{equation}
where $k^2(\varphi)=A^{-2}\left[1-6\beta^2\right]$. From here one can straightforwardly see that the self-interactions of gravity will be modified with respect to GR due to the presence of the conformal coupling $A^{-2}$ in front of the Ricci Scalar. 
In local environments we evaluate it on the screened solution to obtain a time variation, at leading order, given by:
\begin{equation}
\frac{\dot{G}_{gw}}{G_{gw}}=  2\frac{\dot{A}_b}{A_b}.
\end{equation}

We notice from these results that there is one main effect that generates a technical difference in the gravitational couplings $G_N$ and $G_{gw}$, which comes from the fact that matter feels a fifth force from the scalar field, which ends up contributing directly to $G_N$.
However, since $G_{gw}$ is the coupling of the gravitational self-interactions, it is not sensitive to an additional fifth force but instead simply to the value of the scalar field itself. 

Regarding the time evolution of these couplings, we explicitly estimate that:
\begin{align}
\frac{\dot{G_N}}{G_N}&=  2\left(\beta\frac{\varphi_b}{M_P}\right)\frac{\dot{\varphi}_b}{\varphi_b}+\left(2\frac{\dot{\varphi}_b}{\varphi_b}-\frac{\dot{m}_\text{eff}}{m_\text{eff}}+\frac{\dot{A}_b}{A_b}\right)[(m_\text{eff}/A_b)r]\frac{F_{5AB}}{F_{AB}}\nonumber \\
& \sim  \left(\beta\frac{\varphi_b}{M_P}\right)T^{-1}+T^{-1}[(m_\text{eff}/A_b)r]\frac{F_{5AB}}{F_{AB}},\label{GNdot}\\
 \frac{\dot{G}_{gw}}{G_{gw}}& =     2\left(\beta\frac{\varphi_b}{M_P}\right)\frac{\dot{\varphi}_b}{\varphi_b}  ,\nonumber \\
 &\sim     \left(\beta\frac{\varphi_b}{M_P}\right)T^{-1} \label{Ggwdot},
\end{align}
where we recall that $T$ is the characteristic timescale at which quantities change in the background, that is, $\dot{a}_\text{eff}/a_\text{eff}\sim T^{-1}$, $\dot{\varphi}_b/\varphi_b\sim T^{-1}$, and $\dot{m}_\text{eff}/m_\text{eff}\sim T^{-1}$. This time $T$ can be taken to be in the Jordan or Einstein frame, as both times are related by $A_b dt=d\tau$ but since $A_b\approx 1$ both times are expected to agree at leading order.
Here we have also used that $|\dot{A}_b/A_b|\sim |(\beta\varphi_b/M_P)|T^{-1}$, and that $\dot{Q}/Q\sim \dot{\varphi}_b/\varphi_b$. From these estimates we see that meanwhile some time evolution in $G_N$ and $G_{gw}$ technically remains, it will be suppressed by the Chameleon screening mechanism as it satisfies $|\beta \varphi_b/M_P|\ll1$, and $|F_{5AB}/F_{AB}|\ll 1$ in local environments in the weak-field regime\footnote{Note that for models with $n<-1$ the background value of the scalar field $\varphi_b$ grows with the energy density $\rho_b$. However, since for stars we nevertheless expect $|\rho_b/M_P^4|\ll1$ for $\beta\sim 1$ then we will still have that $|\beta \varphi_b/M_P|\ll1$ for our cases of interest.}.

For a heavy enough scalar field, then the first term in eq.~(\ref{GNdot}) dominates the overall time variation of $G_N$ (since $F_5$ scales as $e^{-(m_\text{eff}/A_b)r}$ and thus $(m_\text{eff}/A_br)F_5$ approaches zero for $|m_\text{eff}/A_br|\gg 1$), in which case both couplings $G_N$ and $G_{gw}$ and their time variations coincide. This happens when screening is very efficient. 
In the regime where the fifth force is highly suppressed, both couplings also coincide in models with Vainshtein screening \cite{Kimura:2011dc} belonging to the class of Horndeski theories \cite{Horndeski:1974wa}. However, for these models the time variation of $G_N$ and $G_{gw}$ is not necessarily screened and may be simply of order $T^{-1}$. This result would be in disagreement with LLR observations that constrain $|\dot{G}_N/G_N|\lesssim 10^{-3}H_0$, and therefore these models would be disfavored unless their parameters are fine tuned to circumvent this constraint (see e.g.~\cite{Barreira:2014jha}). For Chameleon models, since the background time variations can be faster than cosmological time variations, we find that LLR data constrain $|\beta\varphi_b/M_P|\ll 10^{-3}$ on galactic environments which, as we will estimate next, will be clearly satisfied in the Solar System.

In general, for $\beta$ of order unity, Chameleon screening will be effective if $|\beta\varphi_b/M_P|\lesssim \Phi_N$ \cite{Burrage:2017qrf}, with $\Phi_N$ being the Newtonian gravitational potential of the source $\delta\rho_\text{eff}$ near its surface. For the Earth, it is estimated that $\Phi_N\sim 10^{-9}$, for the Moon $\Phi_N\sim 10^{-11}$, and for the Sun $\Phi_N\sim 10^{-6}$. Viable models should then always have at least $|\beta\varphi_b/M_P|\lesssim 10^{-6}$. 
More specifically, Solar System and Laboratory tests \cite{Khoury:2003rn} on fifth forces indicate that for a galactic background with $\rho_b\sim  10^{-24}$gr/cm${}^3$, then $|\beta\varphi_b/M_P|\lesssim 10^{-19}$ for $\beta$ and $n$ order unity\footnote{This estimate is obtained by assuming $M\lesssim 10^{-3}$eV and that $\varphi_b$ is obtained minimizing the effective potential $V_\text{eff}(\varphi_b,\rho_b)$, with $\beta$ and $n$ order unity in eq.~(\ref{Potential}).} and hence viable Chameleon models certainly satisfy LLR constraints on time variations of $G_N$, even if $T$ is many orders of magnitude faster than cosmological timescales. 

For the case of $G_{gw}$, the predictions of Chameleon models will lie below the sensitivity that third-generation GW detectors will reach \cite{Belgacem:2019pkk,Nishizawa:2019rra, DAgostino:2019hvh, Bonilla:2019mbm} and will therefore become effectively indistinguishable from $\Lambda$CDM for those observables. We note that time varying gravitational couplings can also be constrained by analyzing the waveform of gravitational waves at the moment of emission. However, those constraints are expected to be rather weak \cite{Carson:2019kkh}.

On the other hand, if the fifth force has a rather long range on galactic backgrounds (i.e.~scalar field is light), as it may actually happen in Solar System scales when $\beta$ and $n$ are both order unity \cite{Khoury:2003rn}, then we find that $|m_\text{eff}r/A_b|\ll 1$. In this case, the two terms in eq.~(\ref{GNdot}) may be comparable depending on the amplitude of the fifth force. Specifically, for the galactic values previously mentioned we obtain $|\beta\varphi_b/M_P|\lesssim 10^{-19}$, $|m_\text{eff}r/A_b|\sim 10^{-3}$ and for the force between the Moon and Earth we get that $F_{5AB}/F_{AB}\sim 10^{-18}$, where we have used that $Q\sim (\varphi_b/M_p)\Phi_N^{-1}$ \cite{Khoury:2003rn, Burrage:2017qrf} for $\beta\sim 1$. We therefore conclude that in this example of long-range force the first term dominates eq.~(\ref{GNdot}) anyway, and both couplings still coincide. We note that the resulting fifth force between the Moon and the Earth was very small because for both sources only a thin shell of mass participates in the interaction and hence we had that $F_{5AB}/F_{AB}\sim Q_\text{Moon}Q_\text{Earth}\ll 1$. In the case of local regions that are unscreened, the second term in eq.~(\ref{GNdot}) will dominate. This may happen for dwarf galaxies \cite{Vikram:2013uba}.

We emphasize that observations of the GW luminosity distance will not only depend on whether the time evolution of $G_{gw}$ is screened in the Solar System, but also on whether this is the case at the moment of emission, that is, around compact objects such as neutron stars and black holes. 
Similarly, the time evolution of $G_N$ for compact objects will also be important as it can affect the GW waveform and can thus be also observationally tested \cite{Carson:2019kkh}. During the inspiral phase of compact objects, the approximation of weak field and slow moving sources is accurate enough to describe the main characteristics of the GW emission. In that regime, the results found in this paper broadly apply and we thus expect the time evolution of gravitational couplings to be suppressed and no observable deviations from General Relativity. Whether this is the case in the dynamical non-relativistic strong-field regime must be simulated numerically and remains to be checked for Chameleon models. So far, analyses on static scenarios have been performed in \cite{Tsujikawa:2009yf, Babichev:2009td, Upadhye:2009kt, Brax:2017wcj}, where it has been found that relativistic compact stars exist in Chameleon models, and that the profile for the scalar field around such stars may exhibit the thin-shell effect or not depending on the stellar density and equation of state. In addition, Chameleon theories have been found to exhibit the same black hole solutions as GR in a stationary vacuum setting \cite{Sotiriou:2011dz}, but it has also been shown that if black holes are modeled more realistically and assumed to be embedded on a cosmologically evolving scalar field or with surrounding matter, then they develop non-trivial profiles that would effectively translate into non-trivial gravitational couplings \cite{Jacobson:1999vr, Horbatsch:2011ye, Berti:2013gfa, Davis:2014tea, Davis:2016avf, Frolov:2017asg}. Also, critical phenomena such as scalarization and superradiance may also lead to modified effective gravitational couplings by allowing non-trivial black hole and star solutions in scalar-tensor models \cite{Cardoso:2013opa, Salgado:1998sg}. 

We note that the solutions presented here are valid for an approximation in which the background is a homogeneous and isotropic patch and there is only one spherically-symmetric source. In realistic situations, there will be additional modifications due to the presence of other nearby bodies such as the Sun. Since Chameleon screening mechanisms rely on the non-linear structure of the potential, then the solutions are not simple linear superpositions, and therefore a full reanalysis must be done to obtain precise realistic constraints (see a discussion in \cite{2009NuPhS.194..230H}). Nevertheless, the calculations made in this paper are expected to describe the broad features of the system.

As previously mentioned, the detection of gravitational waves can be used to obtain an estimate on the GW luminosity distance, and in eq.~(\ref{dgw}) all the modifications have been attributed to a time-varying $G_{gw}$. However, the dark energy field will also have oscillations that will interact with GW in a non-trivial way. This can induce an exchange of energy between GW and dark energy during the propagation on a non-homogeneous background (see \cite{Dalang:2019rke, Garoffolo:2019mna}), affecting the overall amplitude detected of GW, from which $d_{gw}$ is derived. In models where the time variation of $G_{gw}$ is expected to be highly suppressed, as the Chameleon models presented here, it will be important to estimate this exchange of energy with dark energy oscillations as they could dominate the overall deviation with respect to GR in $d_{gw}$ (albeit this effect may also be small and unobservable). Whether this is the case will be left for future work (see \cite{Garoffolo:2019mna} for a list of the different effects that may contribute to $d_{gw}$ when non-minimal interactions between gravity and a scalar field are present).

\section{Summary}\label{Sec:discussion}

In this paper we have considered a simple model where dark energy is dynamical and has a conformal coupling to gravity. In this scenario, we explicitly work out the time evolution of the system in a local, dense region where the Chameleon screening mechanism occurs in the weak-field non-relativistic regime. Since this model breaks the equivalence principle, there is no universal gravitational coupling anymore and the gravitational strength will depend on the particular bodies participating in the interaction. We calculate the time variation of both $G_N$, the coupling determining the matter-matter gravitational interactions, as well as $G_{gw}$, the coupling determining the gravitational self-interactions of gravity. These couplings are in principle probed by different observations: Lunar Laser Ranging for $G_N$, and gravitational waves for $G_{gw}$. 
We find that both of these couplings are generically different due to the fact that the dark energy field mediates a fifth force that contributes directly to $G_N$ but not $G_{gw}$. In the case of the Solar System embedded in the Milky Way, we find that the fifth force is screened enough that both couplings actually coincide.
We also find that any possible time evolution (induced by evolutions of the galatic background where the local screened region is embedded in) of these couplings is highly suppressed by the Chameleon mechanism, contrary to some models with Vainshtein screening. Therefore, we generically expect their time variations to lie below the sensitivity of LLR and GW observations, and hence this model will be indistinguishable from $\Lambda$CDM in this regard. We note however that since GW observations are concerned with the value of $G_{gw}$ and $G_N$ both at the moment of emission and detection, it is important to also study screening in highly dynamical strong-field regimes to find out the observational effects of the dark energy field near the merger of compact objects, such as black holes and neutron stars. This must be done with numerical simulations and remains to be studied. 

The model presented in this paper shows explicitly why and how the time evolution of both gravitational couplings may differ. Due to the similarities between models with Chameleon and Symmetron and Dilaton screeening (in which screening works by suppressing the interaction between matter and the scalar field in dense environments) \cite{Hinterbichler:2010es, Hinterbichler:2011ca, Brax:2010gi}, we expect the same overall results to hold in all these models. 

Finally, we mention that more general models may allow for observable time variations of the gravitational couplings. For instance, DHOST theories rely on the Vainshtein screening mechanism, where it has been found that the model propagating luminal GW has three free functions $\{A, \alpha_B, \beta\}$ \cite{Creminelli:2017sry, Langlois:2017dyl} that depend on the cosmological time evolution. The Newtonian gravitational coupling is given by $G_N=GA^2(1-\beta)$ \cite{Crisostomi:2019yfo, Hirano:2019scf}. Here, $A$ plays an analogous role to the Chameleon model, that is, it is a conformal factor as in eq.~(\ref{ActionJ}) but it can depend on the value of the scalar field as well as its derivatives. Meanwhile, $G_{gw}$ on a cosmological background is simply given by $G_{gw}=GA^2$ \cite{Langlois:2017mxy, Creminelli:2017sry,  Langlois:2017dyl}. In this model, the difference between $G_N$ and $G_{gw}$ happens because the scalar field contributes non-linearly to the Newtonian gravitational force with the same standard scaling $F\propto 1/r^2$, which results in an additional factor of $(1-\beta)$ in $G_N$. In other words, the fifth force does make non-negligible contributions to the gravitational force but they do not modify the $1/r^2$ shape at leading order.
Assuming that the expression for $G_{gw}$ is also valid in local regions when neglecting the interactions with the scalar field, one may have a model where $A$ is allowed to vary on cosmological timescales (and hence $\dot{G}_{gw}/G_{gw}$ could be falsified with future GW observations), whereas a tuning between $A$ and $\beta$ would be required to suppress the time evolution of $G_N$ to conform to LLR observations. The work in \cite{Crisostomi:2017pjs} illustrates the time-evolution of $G_N$ and $G_{gw}/G_N$ in a sub-class of DHOST models with self-acceleration (no added cosmological constant) and stable cosmological perturbations \cite{Crisostomi:2018bsp}.
Whether models like this remain viable when contrasted against other observations will be left for future work.

%--------------------------------------------------------------------------------------------------------------------------------------------------------

\begin{acknowledgments}
We thank Wayne Hu for useful comments. 
ML and HZ acknowledge support from the Kavli Institute for Cosmological Physics at the University of Chicago through an endowment from the Kavli Foundation and its founder Fred Kavli. 
\end{acknowledgments}

%--------------------------------------------------------------------------------------------------------------------------------------------------------

%\appendix

%\section{Generalised Proca theory}\label{App:GeneralProca}

\bibliographystyle{apsrev4-1}
\bibliography{RefModifiedGravity}

\end{document}